\documentclass[runningheads]{llncs}
\usepackage[T1]{fontenc}
\usepackage{graphicx}
\usepackage{amsmath}  
\usepackage{amssymb}  
\usepackage{bm}       

\usepackage{cite}

\usepackage{booktabs}
\begin{document}
\title{ResWCAE: Biometric Pattern Image Denoising \\ Using Residual Wavelet-Conditioned Autoencoder}
%
%
\author{Youzhi Liang\inst{1} \and
Wen Liang\inst{2}\orcidID{0009-0006-5646-2214}}
\authorrunning{Y. Liang et al.}
\titlerunning{ResWCAE: Biometric Pattern Image Denoising}

%
\institute{Stanford University, Stanford CA 94305, USA \email{youzhil@stanford.edu}\and
Google Research, Mountain View CA 94043, USA \email{liangwen@google.com}}
\maketitle              
\begin{abstract}

The utilization of biometric authentication with pattern images is increasingly popular in compact Internet of Things (IoT) devices. However, the reliability of such systems can be compromised by image quality issues, particularly in the presence of high levels of noise. While state-of-the-art deep learning algorithms designed for generic image denoising have shown promise, their large number of parameters and lack of optimization for unique biometric pattern retrieval make them unsuitable for these devices and scenarios. In response to these challenges, this paper proposes a lightweight and robust deep learning architecture, the Residual Wavelet-Conditioned Convolutional Autoencoder (Res-WCAE) with a Kullback-Leibler divergence (KLD) regularization, designed specifically for fingerprint image denoising. Res-WCAE comprises two encoders - an image encoder and a wavelet encoder - and one decoder. Residual connections between the image encoder and decoder are leveraged to preserve fine-grained spatial features, where the bottleneck layer conditioned on the compressed representation of features obtained from the wavelet encoder using approximation and detail subimages in the wavelet-transform domain. The effectiveness of Res-WCAE is evaluated against several state-of-the-art denoising methods, and the experimental results demonstrate that Res-WCAE outperforms these methods, particularly for heavily degraded fingerprint images in the presence of high levels of noise. Overall, Res-WCAE shows promise as a solution to the challenges faced by biometric authentication systems in compact IoT devices.

\keywords{Computer Vision  \and Signal Processing \and Image Denoising \and Deep Learning for Biometrics.}
\end{abstract}

\section{Introduction}

Biometric authentication has gained popularity with the recent advances in sensing systems and vision algorithms~\cite{weaver2006biometric}. Biometric traits, including voice, gait, face, and fingerprints, are widely used to unlock devices, including phones, computers, door locks, and other Internet of Things (IoT) devices~\cite{mgaga2019review}. Fingerprint recognition, in particular, is an active research area due to the unique and stable ridge and minutiae features of fingerprints. However, fingerprint images are susceptible to image quality issues induced by various impression conditions, such as humidity, wetness, dirt, as well as user behavior, resulting in low-quality images that require noise reduction or inpainting of adjacent areas~\cite{drahansky2017challenges}.  High levels of noise caused by the failure or degradation of sensors and by the parched or drenched conditions of user fingers, can lead to repeated fingerprint recognition failure, therefore necessitating denoising of fingerprint images as a pre-processing step to facilitate subsequent operations such as fingerprint authentication and verification~\cite{dharavath2013study}.

Desnoising of fingerprint images is considered as a subfield of image denoising, a well-established and actively researched area in low-level vision, as it is a crucial step in numerous practical applications, including medical imaging, surveillance and photography~\cite{fan2019brief, tian2020deep}. The fundamental objective of image denoising is to retrieve a noise-free image from a noisy observation image. 
Let $\bm{I}(m,n)$ be a matrix of size $M \times N$, representing a noise-free image, where $(m, n)$ denote pixel coordinates, and let $\bm{J}(m, n)$ be a matrix of the same size representing the corresponding noisy image. According to the image degradation model, the noisy image $\bm{J}(m,n)$ can be described as: $\bm{J}(m, n) = \bm{I}(m, n) + \bm{\epsilon}(m, n)$, where $\bm{\epsilon}(m, n) \sim \mathcal{N}(0, \sigma^2)$ represents the additive noise. This noise follows an additive white Gaussian noise (AWGN) model, which introduces independent Gaussian noise to each pixel in the image. The AWGN model is widely used in image processing as it provides a simple yet effective method to simulate noise and evaluate the performance of denoising algorithms. Importantly, the noise $\bm{\epsilon}(m, n)$ added to each pixel is independent of the noise added to other pixels, making it suitable for testing the robustness of denoising techniques across diverse noise patterns.

State-of-the-art deep learning algorithms have achieved remarkable performance on generic image denoising tasks~\cite{zhang2018ffdnet, zhang2017beyond}. Nevertheless, the associated large number of parameters, ranging from millions to billions, renders them unsuitable for deployment on compact Internet of Things (IoT) devices. Furthermore, these algorithms are designed to denoise images acquired using CMOS sensing with low levels of noise, i.e., $\sigma \leq 50$, which may not be appropriate for fingerprint images acquired using capacitive sensing, in particular under the condition of high levels of noise when hands are too dry or wet, i.e., $\sigma \in [100, 200]$. Fingerprint images are characterized by unique features such as ridges and valleys that must be preserved during the denoising process. Thus, specialized algorithms are required for fingerprint image denoising, which are specifically designed to handle these distinct features targeted for high noisy conditions. Such algorithms leverage fingerprint-specific information, such as ridge orientation or minutiae, to guide the denoising process and protect the critical features of the fingerprint. The use of a generic image denoising algorithm for fingerprint images may not be suitable, as it may result in the loss of vital fingerprint information, reducing the accuracy and reliability of fingerprint recognition systems. 
Traditional fingerprint denoising techniques often struggle with high noise levels, particularly those introduced by adverse environmental conditions and low-quality sensors typical of compact devices~\cite{bae2020fingerprint, prabhu2019u, adiga2019fpd, reddy2008fingerprint}. For instance, methods such as U-Finger model~\cite{prabhu2019u} and Fpd-m-net~\cite{adiga2019fpd}, while effective under controlled conditions, do not perform adequately when noise levels $\sigma$ exceed 50, a common scenario in IoT applications. Furthermore, the computational demands of advanced denoising algorithms, which often require substantial processing power, are impractical for IoT devices with limited hardware capabilities. To address these challenges, our research introduces the Residual Wavelet-Conditioned Convolutional Autoencoder (ResWCAE), a novel architecture that combines the robustness of wavelet-based feature extraction with the efficiency of deep learning models. This approach not only manages the high variability in noise levels effectively but also adheres to the computational constraints of IoT devices. ResWCAE leverages a dual-encoder system that incorporates both traditional image encoding and wavelet transform-based encoding, enabling it to capture and preserve the intricate patterns of fingerprints that are crucial for authentication purposes, even in heavily degraded images. 

In this paper, we propose and evaluate a deep learning architecture, a Residual Wavelet-Conditioned Convolutional Autoencoder (Res-WCAE), to retrieve the underlying intricate and unique fingerprint features, dedicately developed for denoising heavily degraded biometric pattern images in the presence of significantly high levels of noise interference for images with capacitive sensing techniques.  We evaluate our model performance using two datasets consisting of AWGN and synthetic images. Our proposed approach exhibits both lightweight, accurate and robust features, thus making it a highly suitable candidate for deployment in practical scenarios, including small Internet of Things (IoT) devices.


\section{Methods}
\begin{figure}[ht]
  \includegraphics[width=130mm]{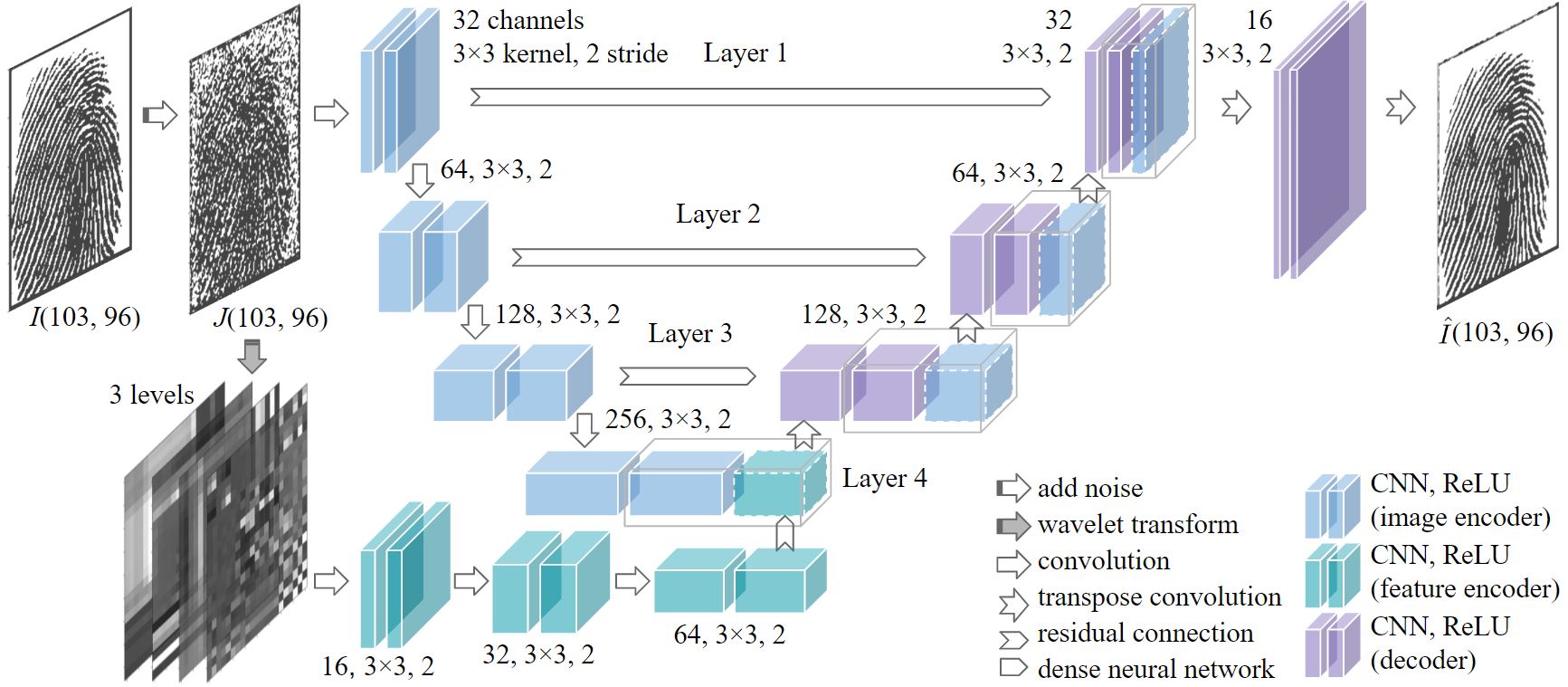}
  \centering
  \caption{A schematic for the architecture of a Residual Wavelet-Conditioned Convolutional Autoencoder (ResWCAE), including a sample noise-free image, ${\bm{I}(103, 96)}$, and a denoised image, $\hat{\bm{I}}(103, 96)$. }
  \label{fig:rescae_architecture}
\end{figure}

Our research proposes a Residual Wavelet-Conditioned Convolutional Autoencoder (Res-WCAE) architecture for capturing fine-grained features in fingerprint pattern images obtained through capacitive sensing devices such as cell phones and other compact Internet of Things (IoT) devices. The Res-WCAE architecture comprises two encoders - an image encoder and a wavelet encoder - and one decoder. These encoders work in unison to construct condition layer for the decoder by leveraging compressed features in both the spatial domain and frequency domain, as illustrated in Fig.~\ref{fig:rescae_architecture}. Additionally, residual connections between the image encoder and decoder have been incorporated to enhance the spatial details of the fingerprint patterns. The Res-WCAE can handle a wide range of noise levels with a standard deviation of $\sigma_{\epsilon}$ ranging from 0 to 200, and achieves state-of-the-art denoising performance, notably including in noise levels that were not covered in prior research, to the best of our knowledge.

\subsection{Image Encoder}
The image encoder consists of four down-sampling convolutional layers, producing the condensed representation of image information as well as supplementing the decoder with fine-grained spatial details through residual connections. The input to the image encoder is a 2D gray-scale image that is passed through a sequence of down-sampling layers. These layers reduce the spatial dimensions of the input image while simultaneously increasing the number of channels. The first layer applies a 3x3 convolution with 32 filters and a stride of 2, followed by a Rectified Linear Unit (ReLU) activation function. The output of this layer is then passed to the second layer, which applies a similar convolution with 64 filters and a stride of 2, also followed by a ReLU activation function. The same process is repeated in the third and fourth down-sampling layers, which have 128 and 256 filters respectively. The output of each layer in the image decoder can be represented as follows:

$$\bm{y}_{\mathcal{E}, img}^{[l]} =  \mathcal{F}_{\mathcal{E}, img}^{[l]} \left( 
\bm{y}_{\mathcal{E},{img}}^{[l-1]}; \mathbf{\Theta}_{\mathcal{E}, img}^{[l]} \right) , $$

where $\bm{y}_{\mathcal{E}, img}^{[l]}$ denotes the output of the image encoder at layer $l$, $\mathcal{F}_{\mathcal{E}, img}^{[l]}(\cdot; \mathbf{\Theta}_{\mathcal{E}, img}^{[l]})$ denotes the function of the convolutional neural network followed by a ReLU activation function at layer $l$, with trainable parameters $\mathbf{\Theta}_{\mathcal{E}, img}^{[l]}$ of the image encoder. The input to the first layer of the image encoder is the noisy image, i.e. $\bm{y}_{\mathcal{E}, img}^{[0]} = \bm{J}(M, N)$.

The output, $\bm{y}_{\mathcal{E}, img}^{[l]}$, of each down-sampling layer in the image decoder not only serves as input for the next down-sampling layer but also partially serves as input for the corresponding upsampling layers in the decoder. To preserve important spatial details of the image, the ResCAE employs residual connections between the image encoder and decoder blocks. These connections allow the network to propagate information from the image encoder to the decoder, while retaining fine-grained spatial details of the image. Specifically, the output of each down-sampling block is concatenated with the output of the corresponding up-sampling block, enabling the network to preserve the spatial information and recover intricate details, which will be elaborated in the decoder section.

\subsection{Wavelet Encoder}

The wavelet encoder employs wavelet transform to extract features from the input image. The wavelet transform coefficients are subsequently passed through a sequence of convolutional layers, which reduce the number of channels while preserving the spatial dimensions, where three convolutional layers with 16, 32, and 64 filters, respectively, followed by a rectified linear unit (ReLU) activation function.

Wavelet transform is a widely-used technique in signal and image processing due to its capability to capture both time and frequency domain information~\cite{wang2022asymptomatic, dong2020x}. In image processing, wavelet transform can extract both high-frequency details and low-frequency approximations, offering a multi-resolution analysis of the image by decomposing the image into several levels of detail, which is highly beneficial for capturing fine-grained features of an image while reducing noise and redundancy~\cite{fan2022high, you2023research}. Wavelet transforms have become integral in image compression for their efficiency in representing and encoding details. Initially, wavelet techniques focused on preserving edge information crucial for maintaining image quality during compression~\cite{chrysafis2000line}. Innovations have included methods to enhance compression efficiency through memory reduction and the introduction of line-based approaches~\cite{he2007peak}. A significant advancement came with the development of the peak transform (PT), a nonlinear geometric transform that optimized image representation and coding. This approach was further refined with a dynamic programming solution that maximized coding gain, leading to the creation of a specialized PT encoder~\cite{kumar2009super}. The application of wavelet transforms extended into super-resolution reconstruction, where high-resolution images compressed at the source could be effectively reconstructed at the decoder from lower-resolution outputs. Techniques have also been developed to preserve edges using wavelet and edge-based segmentation, achieving high compression ratios while maintaining high image quality~\cite{naveen2019lossless}. Recent developments have seen the integration of convolutional neural networks with wavelet-like transforms to address the complexities of compressing natural images, which do not always conform to traditional wavelet assumptions. Additionally, novel algorithms for image blind deblurring have utilized deep multi-level wavelet transforms to enhance image quality by leveraging advanced features such as multi-scale blocks and feature fusion techniques~\cite{ma2019iwave}. Furthermore, wavelet-based deep auto encoder-decoder networks have been tailored to manage various frequency components within images, showing marked improvements in compression metrics such as PSNR and SSIM~\cite{mishra2020wavelet}. Overall, wavelet transform-based techniques continue to evolve, offering promising solutions for efficient image compression that retains essential features like edges and details. Ongoing research aims to further refine and innovate within this field to enhance the performance of image compression technologies.

The two-dimensional (2D) wavelet decomposition of a discrete image $\bm{J}(M,N)$ into $K$ octaves results in $3K+1$ subimages that represent the image at different scales and orientations:

$$\mathbb{J}_K = \left[ \bm{J}_K, \bigcup_{k=1}^K \{\bm{j}_k^1, \bm{j}_k^2, \bm{j}_k^3 \} \right],$$ 

where $\bm{J}_K$ denotes a low-resolution approximation of the original image $\bm{J}(M, N)$ and $\{\bm{j}_k^1, \bm{j}_k^2, \bm{j}_k^3 \}$ represents the wavelet subimages containing the image details at different scales ($2^k$) and orientations.

Fingerprints exhibit quasi-periodic patterns with dominant frequencies typically located in the middle frequency channels of the wavelet decomposition, as noted in prior research \cite{tico2001wavelet, le2020fingerprint}. By taking into account ridge orientation and spatial frequency across different regions of the image, one can better capture the inherent nature of the fingerprint image~\cite{hu2021procedure, joshi2021discrete}. We employ a three-layer convolutional neural network (CNN) to extract the condensed feature representation in the wavelet-transform domain. The output of each layer in the wavelet encoder can be represented as follows:

$$\bm{y}_{\mathcal{E}, wvl}^{[l]} =  \mathcal{F}_{\mathcal{E}, wvl}^{[l]} \left( 
\bm{y}_{\mathcal{E},{wvl}}^{[l-1]}; \mathbf{\Theta}_{\mathcal{E}, wvl}^{[l]} \right) , $$

where $\bm{y}_{\mathcal{E}, wvl}^{[l]}$ denotes the output of the wavelet encoder at layer $l$, $\mathcal{F}_{\mathcal{E}, wvl}^{[l]}(\cdot; \mathbf{\Theta}_{\mathcal{E}, wvl}^{[l]})$ denotes the function of the convolutional neural network followed by a ReLU activation at layer $l$ with trainable parameters $\mathbf{\Theta}_{\mathcal{E}, wvl}^{[l]}$. The input to the first wavelet encoder layer is the 2D wavelet decomposition subimages $\mathbb{J}_K$ . We leverage the subimages of wavelet coefficients $\mathbb{J}_K$ with a level of three obtained using Symlets wavelet as input to a three-layer CNN, enabling us to extract a condensed representation of the wavelet transform domain features, as demonstrated in previous works~\cite{sridhar2014wavelet, chen2013wavelet}. Our wavelet encoder is designed to construct an adaptive trainable and parametrized thresholding technique, in contrast to the soft and hard thresholding techniques that are widely used in prior literature~\cite{mgaga2019review, golilarz2017image, zhao2015improved}.

\subsection{Decoder}
The decoder of the network consists of a sequence of up-sampling layers that progressively increase the spatial dimensions of the input while decreasing the number of channels. The up-sampling process initiates with a 3x3 transpose convolution that employs 128 filters and a stride of 2, followed by a rectified linear unit (ReLU) activation function. This process is repeated in the next two layers, with 64 and 32 filters respectively. Finally, the last layer applies a 3x3 transpose convolution with a single filter and a sigmoid activation function to produce the gray-scale image.

The condition layer incorporates the compressed representation of the image and concatenates it with the adaptive compressed representation of the wavelet domain features. By integrating the conditional input, the decoder reconstructs data that is specific to the fingerprint scenario. The output of the encoder layer that takes the condition layer as input can be expressed as:

$$\bm{y}_{\mathcal{D}}^{[3]} =  \mathcal{F}_{\mathcal{D}}^{[3]} \left( \left[
\bm{y}_{\mathcal{E},{img}}^{[4]} \parallel \bm{y}_{\mathcal{E},{wvl}}^{[3]} \right]; \mathbf{\Theta}_{\mathcal{D}}^{[3]} \right),$$

where $\bm{y}_{\mathcal{D}}^{[3]}$ denotes the output of the decoder at layer 3, $\mathcal{F}_{\mathcal{D}}^{[3]}(\cdot; \mathbf{\Theta}_{\mathcal{D}}^{[3]})$ denotes the function of the convolutional neural network followed by a ReLU activation at layer 3 with trainable parameters $\mathbf{\Theta}_{\mathcal{D}}^{[3]}$ and $[ \cdot \parallel \cdot ]$ denotes the concatenation operation. The output of the last decoder layer is the denoised image $\hat{\bm{I}}(M,N) = \bm{y}_{\mathcal{D}}^{[L]}$.


In addition, the network utilizes residual connections between the image encoder and decoder blocks to enhance performance on intricate spatial details~\cite{zini2020deep, chatterjee2022improving}. These connections enable the network to transmit information from the image encoder to the decoder, while retaining crucial fingerprint image details. In particular, the output of each down-sampling block is concatenated with the output of the corresponding up-sampling block, which helps maintain spatial information and facilitates the network's ability to recover fine-grained details. The final decoder layer incorporates a bilinear interpolation technique to upsample the feature maps in the decoder blocks, thereby restoring the spatial resolution of the image. The output of each layer in the decoder can be represented as follows:

$$\bm{y}_{\mathcal{D}}^{[l]} =  \mathcal{F}_{\mathcal{D}}^{[l]} \left( \left[
\bm{y}_{\mathcal{D}}^{[l + 1]} \parallel \bm{y}_{\mathcal{E},{wvl}}^{[l+1]} \right] ; \mathbf{\Theta}_{\mathcal{D}}^{[l]} \right) . $$

To enhance the generalizability of our model, we introduce a regularized cost function that incorporates Kullback–Leibler (KL) divergence regularization through the use of a prior distribution~\cite{kingma2015variational, marojevic2018measuring}. The regularized cost function for Res-WCAE is formulated as the expected loss over the training set using the $L^2$-norm, along with the expected loss over the model parameters using KL divergence, expressed as follows:

$$ \mathcal{L} \left( \mathbf{\Theta} \right) = \mathbb{E}_{\bm{I}} \left\| \bm{y}_{\mathcal{D}}^{[L]} - {\bm{I}} \right\|^2 + \lambda \mathbb{E}_{\bm{y}} D_{\text{KL}} \left( \bm{y}_{\mathcal{D}}^{[L]} \parallel \bm{I} \right) $$

where $\mathbf{\Theta}$ denotes all the trainable parameters, $\lambda D_{\text{KL}} \left( \bm{y}_{\mathcal{D}}^{[L]} \parallel \bm{I} \right)$ denotes the KL divergence of $\bm{y}_{\mathcal{D}}^{[L]}$ from prior distribution $\bm{I}$. The inclusion of KL divergence regularization in our model aims to prevent overfitting to a single training instance, analogous to adapting the target distribution performed by conventional backpropagation algorithms~\cite{yu2013kl}. KL divergence regularization is a versatile tool in machine learning, effectively enhancing model performance across a range of applications. It has been used to adapt DNN acoustic models by focusing on training the context-independent parts of the network~\cite{toth2016adaptation}. This technique has shown substantial benefits in conversational model adaptation by incorporating information from related source domains~\cite{abraham2017transfer}. In the realm of automatic sleep staging, a personalized approach using KL divergence regularization has addressed challenges associated with training on limited data, significantly reducing the risk of overfitting and maintaining performance close to subject-independent models~\cite{li2018conversational}. Additionally, in exploring alternatives to self-supervised learning, data augmentations have been used instead of KL divergence to promote domain-specific invariances in autoencoder models, demonstrating the adaptability of regularization approaches~\cite{phan2020personalized}. A novel approach involving probabilistic density gaps has also been introduced to overcome some limitations of traditional KL divergence in variational autoencoders, refining their optimization processes~\cite{phan2020personalized}. Overall, KL divergence regularization continues to prove its broad applicability and effectiveness in enhancing model adaptability and preventing overfitting across diverse domains~\cite{tang2024generalized, falcon2021aasae}.

\section{Experiments, Results and Discussion}

\begin{figure}[ht]
  \includegraphics[width=120mm]{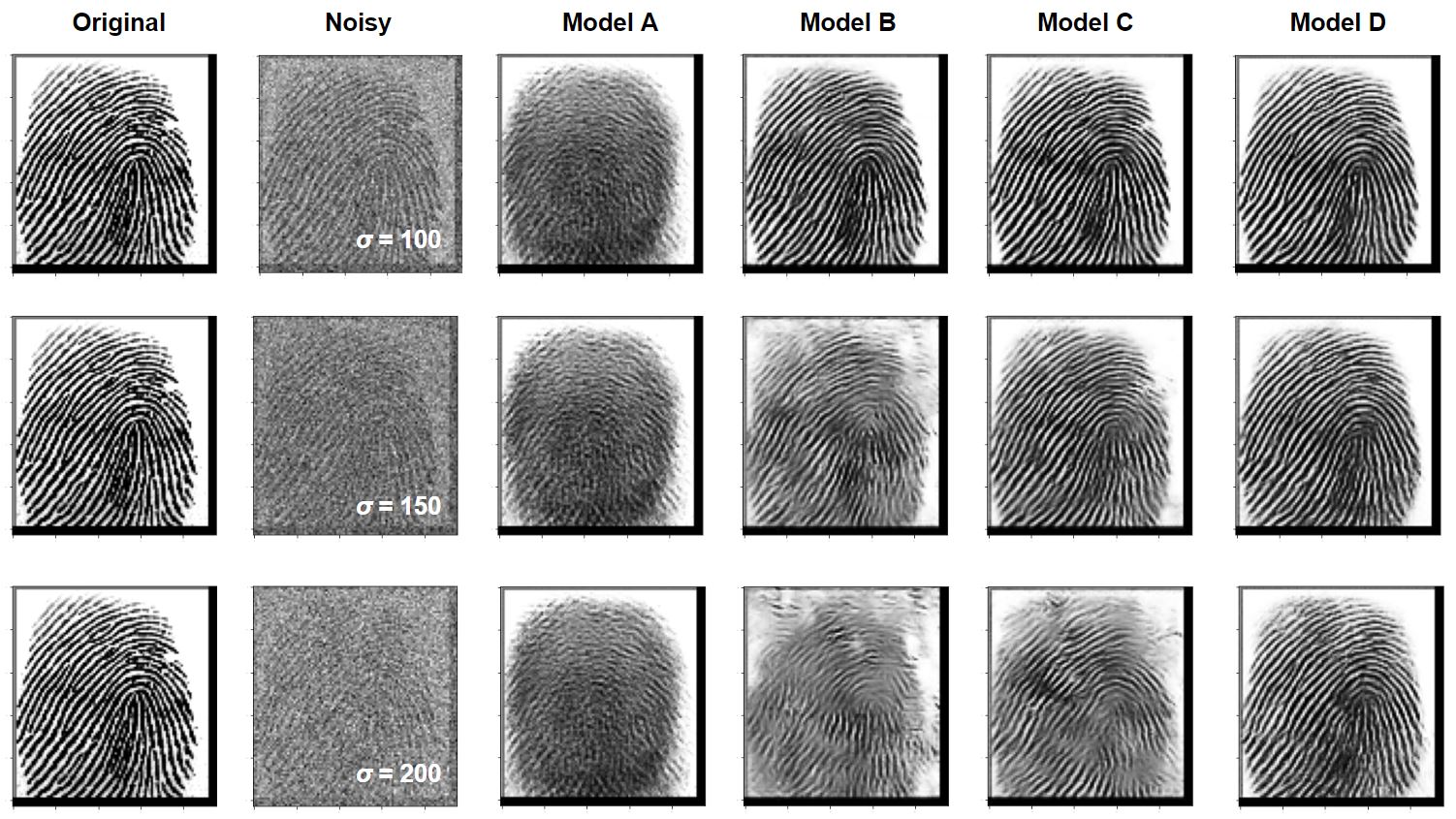}
  \centering
  \caption{Samples of original figures, noisy figures and denoised figures for noise level $\sigma$ from 100 to 200. Model A: AWGN, Model B: Dense NN, Model C: Autoencoder, Model D: Res-WCAE.}
  \label{fig:denoising_comparison}
\end{figure}

The Sokoto Coventry Fingerprint Dataset (SOCOFing) was selected for the purpose of constructing and evaluating the models in this study. SOCOFing is a biometric fingerprint database that has been specifically designed for academic research purposes, as documented in~\cite{shehu2018sokoto}. The dataset is comprised of a total of 6,000 fingerprint images that were collected from 600 African subjects, as outlined in~\cite{shehu2018sokoto}. Figure~\ref{fig:denoising_comparison}(b) provides representative samples from the dataset. During the preprocessing stage, the images in the dataset were converted into grayscale images with a resolution of 103x96 pixels, as per standard practice. All images were originally stored in the .BMP format.

To ensure a reliable evaluation of our models, we partitioned the dataset into training, holdout validation, and testing sets in a 70:15:15 ratio. We initialized the weights and trained all neural network architectures from scratch using a mini-batch size of 32. The learning rate was set to 0.001 and the models were trained for a maximum of 200 iterations. In total, we trained and evaluated four neural network architectures, including dense neural network, Autoencoder, wavelet feature conditioned Autoencoder, and Res-WCAE. We selected models based on their performance on the validation set. Our findings show that Res-WCAE outperformed all other models and achieved state-of-the-art performance in the presence of all levels of noise. Figure~\ref{fig:denoising_comparison} (a) depicts the loss vs epoch, which showed moderate fluctuations due to the mini-batch training. We also evaluated the improved Peak Signal-to-Noise Ratio ($\Delta \textit{PSNR}$) relative to the PSNR of the noisy image ($\bm{J}(m, n)$). As illustrated in Figure~\ref{fig:denoising_comparison} (a) inset, the averaged improved PSNR was approximately 7.5 dB for a wide range of noise levels.

Figure~\ref{fig:denoising_comparison} effectively demonstrates the superior capabilities of the Residual Wavelet-Conditioned Convolutional Autoencoder (ResWCAE) in the denoising and restoration of fingerprint images, particularly under challenging noise conditions. The images displayed reveal that, even with noise levels significantly elevated ($\sigma$ ranging from 100 to 200), the ResWCAE model is adept at reconstructing the intricate features of fingerprint patterns. This includes the minutiae—fine points which are crucial for accurate fingerprint analysis and recognition. The clarity with which these minutiae are preserved is noteworthy, as these features are often lost or obscured when subjected to high levels of noise. The ability of our model to maintain the visibility of critical details such as ridge bifurcation and ridge endings under such adverse conditions highlights its robustness and effectiveness. This capability is particularly important given the typical application of fingerprint technology in security systems, where precision is paramount. The model's performance not only meets but exceeds the benchmark set by previous studies, which typically have examined noise levels only up to $\sigma$=100. By pushing the boundary to $\sigma$=200, we provide compelling evidence of the model’s robustness across a spectrum of real-world scenarios where noise can be a significant impediment. Additionally, the side-by-side comparison of original, noisy, and denoised images within the figure provides a clear visual reference that helps to quantify the improvement in image quality achieved by our model. These comparative visuals not only validate our quantitative metrics but also offer a straightforward, qualitative assessment that can be easily interpreted by practitioners and researchers alike. By expanding the test range beyond traditional studies, our work addresses a critical gap in the existing literature on fingerprint denoising, presenting a solution that is both more versatile and more applicable to real-world conditions. This enhanced demonstration of the model’s capabilities underscores its potential utility in practical applications, where such robustness against high noise levels is crucial.

\begin{figure}[ht]
  \includegraphics[width=120mm]{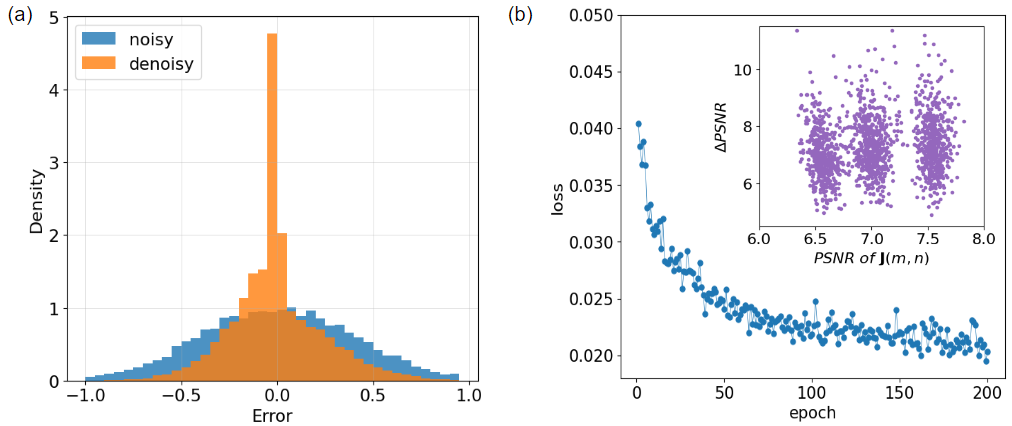}
  \centering
  \caption{(a) Comparative Analysis of Error Distributions for Noisy and Denoised Images (b) Correlation Between Epoch-Based Training Loss and Peak Signal-to-Noise Ratio Enhancements(inset) Scatter Plot of Incremental Peak Signal-to-Noise Ratio Enhancements Versus PSNR of Denoised Images Over Training Epochs}
  \label{fig:training_result}
\end{figure}

Fig.~\ref{fig:training_result} presents a comprehensive analysis of a denoising model's performance. Panel (a) depicts the error distribution for pixel intensity values in both noisy and denoised images, using histograms. In this panel, the x-axis represents the error magnitude and the y-axis the density of errors. The noisy image errors (in blue) display a broader spread, indicating higher variability, while the denoised image errors (in orange) are tightly concentrated around zero, highlighting an enhancement in image accuracy after denoising. Panel (b) illustrates the model's training loss over epochs (on the x-axis against the loss on the y-axis), showing a progressive decline in loss, which signifies improvement in the model's performance. The inset within Panel (b) further explores this relationship by plotting the change in Peak Signal-to-Noise Ratio ($\Delta \text{PSNR}$) against the PSNR of the denoised image ($\text{PSNR of } J(m, n)$), where each point represents an epoch. The trend indicates that an increase in PSNR correlates with a rise in $\Delta \text{PSNR}$, consistent with the decreasing trend in training loss.

We conduct a more rigorous assessment of the denoising models using three evaluation metrics, including Peak Signal-to-Noise Ratio (PSNR), Structural Similarity Index Measure (SSIM), and Mean Squared Error (MSE). The models under study were a Res-WCAE, an autoencoder, and an added noise figure, which served as a baseline for comparison. The results of the study are summarized in Table~\ref{tab:performance_evaluation}. 

\begin{table}[ht]
\centering
\caption{Evaluation of Res-WCAE for $\sigma = 100$}
\begin{tabular}{lcccccc}
\toprule
Dataset & Metric & AWGN & Dense NN & Autoencoder & Res-WCAE\\
\midrule
\cite{shehu2018sokoto} & PSNR & 7.92 & 10.03 & 12.77 & 17.88\\
\cite{shehu2018sokoto} & SSIM & 0.45 & 0.35 & 0.69 & 0.79\\
\cite{shehu2018sokoto} & MSE & 0.17 & 0.12 & 0.06 & 0.02\\

\bottomrule
\end{tabular}
\label{tab:performance_evaluation}
\end{table}

\begin{table}[ht]
\centering
\caption{Evaluation of Res-WCAE using NFIQ}
\begin{tabular}{lcccccc}
\toprule
Dataset & Metric & Autoencoder & Res-WCAE & Data \\
\midrule
\cite{escalera2019chalearn} & NFIQ & 39.5 & 43.3 & 44.7\\
\cite{maio2002fvc2000} & NFIQ & 39.9 & 43.1 & 45.5 \\
new & NFIQ & 35.2 & 39.5 & 37.3\\

\bottomrule
\end{tabular}
\label{tab:performance_evaluation2}
\end{table}

We conducted a detailed evaluation of various denoising models using three key metrics: Peak Signal-to-Noise Ratio (PSNR), Structural Similarity Index Measure (SSIM), and Mean Squared Error (MSE). Models evaluated included a Res-WCAE, an autoencoder, and an Additive White Gaussian Noise (AWGN) model, which served as baseline comparisons. The Res-WCAE was chosen for its sophisticated approach to handling complex noise patterns, while the autoencoder was selected for its general utility in image denoising, and the AWGN model was included as a basic benchmark to highlight the improvements from more advanced architectures. According to Table 1, the Res-WCAE markedly outperforms the others with the highest PSNR of 17.88, highest SSIM of 0.79, and lowest MSE of 0.02, demonstrating its superior noise reduction and image integrity preservation. In contrast, the Dense NN and autoencoder models exhibited lower performance due to their simpler architectures, which are less effective at modeling and removing more complex noise types found in practical applications. Specifically, the Dense NN's lower SSIM of 0.35 and higher MSE of 0.12 indicate a less accurate structural retention and higher error rates, while the autoencoder's intermediate performance (SSIM of 0.69 and MSE of 0.06) suggests moderate effectiveness that does not fully capture or remove noise variances as efficiently as the Res-WCAE. ResWCAE model also outperforms the state-of-the-art models applied in the field of fingerprint pattern denoising, including U-Finger model~\cite{prabhu2019u} and Fpd-m-net~\cite{adiga2019fpd}. 

We further verified our model using three additional datasets. The FVC2000 is the First International Competition for Fingerprint Verification Algorithms. It includes a database of 880 fingerprints, with 110 unique fingers (width) and 8 impressions per finger (depth). The ChaLearn Looking At People dataset consists of 168,000 fingerprint images, with 84,000 ground-truth images generated using the Anguli: Synthetic Fingerprint Generator and 84,000 degraded versions created by applying random artifacts like blur and scratches. In addition, we created a new fingerprint dataset for 100 fingerprints, representing the scenarios in presence of high level of noises. Table~\ref{tab:performance_evaluation2} illustrates the performance evaluation of the Res-WCAE across another three different datasets: two established datasets from ChaLearn \cite{escalera2019chalearn} and Fingerprint Verification Competition \cite{maio2002fvc2000}, and a new dataset introduced in this study. The evaluation focuses on the NFIQ metric, a standard measure for assessing image quality in biometric applications. The NIST Fingerprint Image Quality (NFIQ) is a widely used metric for evaluating the quality of fingerprint images, particularly in biometric systems. It assesses how well a fingerprint image can be used for reliable matching and recognition. For fingerprint image denoising, NFIQ serves as a key indicator of how effectively noise has been reduced, ensuring that the denoised image retains sufficient detail and clarity for accurate fingerprint authentication. A higher NFIQ score generally indicates better image quality, making it an essential benchmark in fingerprint denoising tasks. For the datasets from ChaLearn \cite{escalera2019chalearn} and Fingerprint Verification Competition \cite{maio2002fvc2000}, the Res-WCAE achieved NFIQ scores of 43.3 and 43.1, respectively. These scores represent a notable improvement over the standard autoencoder, which scored 39.5 and 39.9 on the same datasets. While the Res-WCAE enhanced the image quality compared to the autoencoder, it did not entirely match the quality of the original data, which had NFIQ scores of 44.7 and 45.5. This suggests that while Res-WCAE effectively mitigates noise and reconstructs higher-quality images, there is still a slight gap compared to the untouched data. In contrast, the new dataset demonstrated a different trend. The Res-WCAE not only outperformed the autoencoder—achieving an NFIQ score of 39.5 versus 35.2—but also surpassed the quality of the original data, which had an NFIQ score of 37.3. This indicates that the Res-WCAE can enhance image quality beyond the original in certain scenarios, possibly due to its ability to learn and correct inherent imperfections in the data. Overall, these results underscore the effectiveness of Res-WCAE in improving image quality, particularly in datasets where the original images are in presence of high noise. The ResWCAE model combines wavelet-based feature extraction with convolutional encoding, offering a more computationally efficient approach to fingerprint denoising. By using wavelet transforms, the model reduces input data complexity and minimizes the number of operations, leading to faster inference times compared to traditional convolutional methods. This makes ResWCAE particularly suitable for IoT devices, where processing power and memory are limited. It is estimated that ResWCAE achieves a 40-70\% reduction in memory usage and offers similar or faster inference times than models like U-Finger and Fpd-m-net. While U-Finger and Fpd-m-net are effective for high noise levels, their architectures require more memory to store intermediate activations and feature maps. In contrast, ResWCAE’s lightweight design and wavelet-based encoding reduce memory demands, making it more practical for resource-constrained environments. This hybrid approach balances efficiency with high performance, making it ideal for real-time or near-real-time applications on compact devices.

\section{Conclusion}

In conclusion, the increasing popularity of biometric authentication in compact Internet of Things (IoT) devices has raised concerns about the reliability of such systems due to image quality issues, especially when dealing with high levels of noise. This paper addresses these challenges by introducing a novel and robust deep learning architecture called Residual Wavelet-Conditioned Convolutional Autoencoder (Res-WCAE) with Kullback-Leibler divergence (KLD) regularization, specifically designed for fingerprint image denoising. By leveraging two encoders - an image encoder and a wavelet encoder - along with residual connections and a compressed representation of features from the wavelet domain, Res-WCAE effectively preserves fine-grained spatial features, outperforming several state-of-the-art denoising methods, especially in heavily degraded fingerprint images with significant noise. The proposed Res-WCAE offers promising solutions for enhancing the reliability of biometric authentication systems in compact IoT devices, presenting a potential breakthrough in the field of image denoising and biometric pattern retrieval.


\bibliographystyle{unsrt}
\bibliography{reference}

\begin{thebibliography}{10}

\bibitem{weaver2006biometric}
Alfred~C Weaver.
\newblock Biometric authentication.
\newblock {\em Computer}, 39(2):96--97, 2006.

\bibitem{mgaga2019review}
Sboniso~Sifiso Mgaga, Nontokozo~Portia Khanyile, and Jules-Raymond Tapamo.
\newblock A review of wavelet transform based techniques for denoising latent fingerprint images.
\newblock {\em 2019 Open Innovations (OI)}, pages 57--62, 2019.

\bibitem{drahansky2017challenges}
Martin Drahansk{\`y}, Ond{\v{r}}ej Kanich, and Eva B{\v{r}}ezinov{\'a}.
\newblock Challenges for fingerprint recognition—spoofing, skin diseases, and environmental effects: Is fingerprint recognition really so reliable and secure?
\newblock {\em Handbook of Biometrics for Forensic Science}, pages 63--83, 2017.

\bibitem{dharavath2013study}
Krishna Dharavath, Fazal~A Talukdar, and Rabul~H Laskar.
\newblock Study on biometric authentication systems, challenges and future trends: A review.
\newblock In {\em 2013 IEEE international conference on computational intelligence and computing research}, pages 1--7. IEEE, 2013.

\bibitem{fan2019brief}
Linwei Fan, Fan Zhang, Hui Fan, and Caiming Zhang.
\newblock Brief review of image denoising techniques.
\newblock {\em Visual Computing for Industry, Biomedicine, and Art}, 2(1):1--12, 2019.

\bibitem{tian2020deep}
Chunwei Tian, Lunke Fei, Wenxian Zheng, Yong Xu, Wangmeng Zuo, and Chia-Wen Lin.
\newblock Deep learning on image denoising: An overview.
\newblock {\em Neural Networks}, 131:251--275, 2020.

\bibitem{zhang2018ffdnet}
Kai Zhang, Wangmeng Zuo, and Lei Zhang.
\newblock Ffdnet: Toward a fast and flexible solution for cnn-based image denoising.
\newblock {\em IEEE Transactions on Image Processing}, 27(9):4608--4622, 2018.

\bibitem{zhang2017beyond}
Kai Zhang, Wangmeng Zuo, Yunjin Chen, Deyu Meng, and Lei Zhang.
\newblock Beyond a gaussian denoiser: Residual learning of deep cnn for image denoising.
\newblock {\em IEEE transactions on image processing}, 26(7):3142--3155, 2017.

\bibitem{bae2020fingerprint}
Jungyoon Bae, Han-Soo Choi, Sujin Kim, and Myungjoo Kang.
\newblock Fingerprint image denoising and inpainting using convolutional neural network.
\newblock {\em Journal of the Korean Society for Industrial and Applied Mathematics}, 24(4):363--374, 2020.

\bibitem{prabhu2019u}
Ramakrishna Prabhu, Xiaojing Yu, Zhangyang Wang, Ding Liu, and Anxiao Jiang.
\newblock U-finger: Multi-scale dilated convolutional network for fingerprint image denoising and inpainting.
\newblock In {\em Inpainting and Denoising Challenges}, pages 45--50. Springer, 2019.

\bibitem{adiga2019fpd}
Sukesh Adiga~V and Jayanthi Sivaswamy.
\newblock Fpd-m-net: Fingerprint image denoising and inpainting using m-net based convolutional neural networks.
\newblock In {\em Inpainting and Denoising Challenges}, pages 51--61. Springer, 2019.

\bibitem{reddy2008fingerprint}
G~Jagadeeswar Reddy, T~Jaya~Chandra Prasad, and MN~Giri Prasad.
\newblock Fingerprint image denoising using curvelet transform.
\newblock {\em ARPN Journal of Engineering and Applied Sciences}, 3(3):31--35, 2008.

\bibitem{wang2022asymptomatic}
Guowei Wang, Shuli Guo, Lina Han, Anil~Baris Cekderi, Xiaowei Song, and Zhilei Zhao.
\newblock Asymptomatic covid-19 ct image denoising method based on wavelet transform combined with improved pso.
\newblock {\em Biomedical Signal Processing and Control}, 76:103707, 2022.

\bibitem{dong2020x}
Hanlei Dong, Liguo Zhao, Yunxing Shu, and Neal~N Xiong.
\newblock X-ray image denoising based on wavelet transform and median filter.
\newblock {\em Applied Mathematics and Nonlinear Sciences}, 5(2):435--442, 2020.

\bibitem{fan2022high}
Lei Fan, Yongjun Wang, Hongxin Zhang, Chao Li, and Xiangjun Xin.
\newblock High-accuracy 3d contour measurement by using the quaternion wavelet transform image denoising technique.
\newblock {\em Electronics}, 11(12):1807, 2022.

\bibitem{you2023research}
Ning You, Libo Han, Daming Zhu, and Weiwei Song.
\newblock Research on image denoising in edge detection based on wavelet transform.
\newblock {\em Applied Sciences}, 13(3):1837, 2023.

\bibitem{chrysafis2000line}
Christos Chrysafis and Antonio Ortega.
\newblock Line-based, reduced memory, wavelet image compression.
\newblock {\em IEEE Transactions on Image processing}, 9(3):378--389, 2000.

\bibitem{he2007peak}
Zhihai He.
\newblock Peak transform-a nonlinear transform for efficient image representation and coding.
\newblock In {\em 2007 IEEE International Conference on Image Processing}, volume~3, pages III--177. IEEE, 2007.

\bibitem{kumar2009super}
CN~Ravi Kumar, VK~Ananthashayana, et~al.
\newblock Super resolution reconstruction of compressed low resolution images using wavelet lifting schemes.
\newblock In {\em 2009 Second International Conference on Computer and Electrical Engineering}, volume~2, pages 629--633. IEEE, 2009.

\bibitem{naveen2019lossless}
R~Naveen~Kumar, BN~Jagadale, and JS~Bhat.
\newblock A lossless image compression algorithm using wavelets and fractional fourier transform.
\newblock {\em SN Applied Sciences}, 1(3):266, 2019.

\bibitem{ma2019iwave}
Haichuan Ma, Dong Liu, Ruiqin Xiong, and Feng Wu.
\newblock iwave: Cnn-based wavelet-like transform for image compression.
\newblock {\em IEEE Transactions on Multimedia}, 22(7):1667--1679, 2019.

\bibitem{mishra2020wavelet}
Dipti Mishra, Satish~Kumar Singh, and Rajat~Kumar Singh.
\newblock Wavelet-based deep auto encoder-decoder (wdaed)-based image compression.
\newblock {\em IEEE Transactions on Circuits and Systems for Video Technology}, 31(4):1452--1462, 2020.

\bibitem{tico2001wavelet}
Marius Tico, Pauli Kuosmanen, and Jukka Saarinen.
\newblock Wavelet domain features for fingerprint recognition.
\newblock {\em Electronics Letters}, 37(1):1, 2001.

\bibitem{le2020fingerprint}
Ngoc~Tuyen Le, Jing-Wein Wang, Duc~Huy Le, Chih-Chiang Wang, and Tu~N Nguyen.
\newblock Fingerprint enhancement based on tensor of wavelet subbands for classification.
\newblock {\em IEEE Access}, 8:6602--6615, 2020.

\bibitem{hu2021procedure}
Zhengbing Hu, Ihor Tereikovskyi, Denys Chernyshev, Liudmyla Tereikovska, Oleh Tereikovskyi, and Dong Wang.
\newblock Procedure for processing biometric parameters based on wavelet transformations.
\newblock {\em International Journal of Modern Education and Computer Science}, 13(2):11--22, 2021.

\bibitem{joshi2021discrete}
Suvarna Joshi.
\newblock Discrete wavelet transform based approach for touchless fingerprint recognition.
\newblock In {\em Proceedings of International Conference on Data Science and Applications: ICDSA 2021, Volume 1}, pages 397--412. Springer, 2021.

\bibitem{sridhar2014wavelet}
Siripurapu Sridhar, P~Rajesh Kumar, and KV~Ramanaiah.
\newblock Wavelet transform techniques for image compression-an evaluation.
\newblock {\em International journal of image, graphics and signal processing}, 6(2):54, 2014.

\bibitem{chen2013wavelet}
Guangyi Chen, Wenfang Xie, and Yongjia Zhao.
\newblock Wavelet-based denoising: A brief review.
\newblock In {\em 2013 fourth international conference on intelligent control and information processing (ICICIP)}, pages 570--574. IEEE, 2013.

\bibitem{golilarz2017image}
Noorbakhsh~Amiri Golilarz and Hasan Demirel.
\newblock Image de-noising using un-decimated wavelet transform (uwt) with soft thresholding technique.
\newblock In {\em 2017 9th International Conference on Computational Intelligence and Communication Networks (CICN)}, pages 16--19. IEEE, 2017.

\bibitem{zhao2015improved}
Rui-Mei Zhao and Hui-min Cui.
\newblock Improved threshold denoising method based on wavelet transform.
\newblock In {\em 2015 7th International Conference on Modelling, Identification and Control (ICMIC)}, pages 1--4. IEEE, 2015.

\bibitem{zini2020deep}
Simone Zini, Simone Bianco, and Raimondo Schettini.
\newblock Deep residual autoencoder for blind universal jpeg restoration.
\newblock {\em IEEE Access}, 8:63283--63294, 2020.

\bibitem{chatterjee2022improving}
Sankhadeep Chatterjee, Asit~Kumar Das, Janmenjoy Nayak, and Danilo Pelusi.
\newblock Improving facial emotion recognition using residual autoencoder coupled affinity based overlapping reduction.
\newblock {\em Mathematics}, 10(3):406, 2022.

\bibitem{kingma2015variational}
Durk~P Kingma, Tim Salimans, and Max Welling.
\newblock Variational dropout and the local reparameterization trick.
\newblock {\em Advances in neural information processing systems}, 28, 2015.

\bibitem{marojevic2018measuring}
Vuk Marojevic, Aditya~V Padaki, Raghunandan~M Rao, and Jeffrey~H Reed.
\newblock Measuring hardware impairments with software-defined radios.
\newblock In {\em 2018 IEEE Frontiers in Education Conference (FIE)}, pages 1--6. IEEE, 2018.

\bibitem{yu2013kl}
Dong Yu, Kaisheng Yao, Hang Su, Gang Li, and Frank Seide.
\newblock Kl-divergence regularized deep neural network adaptation for improved large vocabulary speech recognition.
\newblock In {\em 2013 IEEE International Conference on Acoustics, Speech and Signal Processing}, pages 7893--7897. IEEE, 2013.

\bibitem{toth2016adaptation}
L{\'a}szl{\'o} T{\'o}th and G{\'a}bor Gosztolya.
\newblock Adaptation of dnn acoustic models using kl-divergence regularization and multi-task training.
\newblock In {\em Speech and Computer: 18th International Conference, SPECOM 2016, Budapest, Hungary, August 23-27, 2016, Proceedings 18}, pages 108--115. Springer, 2016.

\bibitem{abraham2017transfer}
Basil Abraham, Tejaswi Seeram, and Srinivasan Umesh.
\newblock Transfer learning and distillation techniques to improve the acoustic modeling of low resource languages.
\newblock In {\em INTERSPEECH}, pages 2158--2162, 2017.

\bibitem{li2018conversational}
Juncen Li, Ping Luo, Fen Lin, and Bo~Chen.
\newblock Conversational model adaptation via kl divergence regularization.
\newblock In {\em Proceedings of the AAAI Conference on Artificial Intelligence}, volume~32, 2018.

\bibitem{phan2020personalized}
Huy Phan, Kaare Mikkelsen, Oliver~Y Ch{\'e}n, Philipp Koch, Alfred Mertins, Preben Kidmose, and Maarten De~Vos.
\newblock Personalized automatic sleep staging with single-night data: a pilot study with kullback--leibler divergence regularization.
\newblock {\em Physiological measurement}, 41(6):064004, 2020.

\bibitem{tang2024generalized}
Yunhao Tang, Zhaohan~Daniel Guo, Zeyu Zheng, Daniele Calandriello, R{\'e}mi Munos, Mark Rowland, Pierre~Harvey Richemond, Michal Valko, Bernardo~{\'A}vila Pires, and Bilal Piot.
\newblock Generalized preference optimization: A unified approach to offline alignment.
\newblock {\em arXiv preprint arXiv:2402.05749}, 2024.

\bibitem{falcon2021aasae}
William Falcon, Ananya~Harsh Jha, Teddy Koker, and Kyunghyun Cho.
\newblock Aasae: Augmentation-augmented stochastic autoencoders.
\newblock {\em arXiv preprint arXiv:2107.12329}, 2021.

\bibitem{shehu2018sokoto}
Yahaya~Isah Shehu, Ariel Ruiz-Garcia, Vasile Palade, and Anne James.
\newblock Sokoto coventry fingerprint dataset.
\newblock {\em arXiv preprint arXiv:1807.10609}, 2018.

\bibitem{escalera2019chalearn}
Sergio Escalera, Mart{\'\i} Soler, Stephane Ayache, Umut G{\"u}{\c{c}}l{\"u}, Jun Wan, Meysam Madadi, Xavier Bar{\'o}, Hugo~Jair Escalante, and Isabelle Guyon.
\newblock Chalearn looking at people: Inpainting and denoising challenges.
\newblock In {\em Inpainting and Denoising Challenges}, pages 23--44. Springer, 2019.

\bibitem{maio2002fvc2000}
Dario Maio, Davide Maltoni, Raffaele Cappelli, James~L. Wayman, and Anil~K. Jain.
\newblock Fvc2000: Fingerprint verification competition.
\newblock {\em IEEE transactions on pattern analysis and machine intelligence}, 24(3):402--412, 2002.

\end{thebibliography}
\end{document}